\documentstyle[a4,epsf,12pt]{article}
\begin{document}
\titlepage
\begin{flushright}
October 1999
\end{flushright}
\vskip 1cm
\begin{center}
{\bf \Large Quintessence and Supergravity}
\end{center}

\vskip 1cm

\begin{center}
{\large Ph. Brax$^{a}$\footnote{email: brax@spht.saclay.cea.fr} $\&$ 
J. Martin$^{b}$\footnote{email: martin@edelweiss.obspm.fr}}
\end{center}
\vskip 0.5cm
\begin{center}
$^a$ {\it Service de Physique Th\'eorique, 
CEA-Saclay\\
F-91191 Gif/Yvette Cedex, France.}\\
$^b$ {\it DARC, Observatoire de Paris-Meudon UMR 8629, \\
92195 Meudon Cedex, France.}
\end{center}

\vskip 2cm
\begin{center}
{\large Abstract}
\end{center}

\noindent
In the context of quintessence, the concept of tracking solutions allows
to address the fine-tuning and coincidence problems. When the field is on tracks
today, one has $Q\approx m_{\rm Pl}$ demonstrating that, generically, any 
realistic model of quintessence must be based on supergravity. We construct 
the most simple model for which the scalar potential is positive. The scalar 
potential deduced from the supergravity model has the 
form $V(Q)=\frac{\Lambda^{4+\alpha}}{Q^{\alpha}}e^{\frac{\kappa}{2}Q^2}$. We show 
that despite the appearence of positive powers 
of the field, the coincidence problem is still solved. If $\alpha \ge 11$, the 
fine-tuning problem can be overcome. Moreover, 
due to the presence of the exponential term, the value of the equation 
of state, $\omega _Q$, is pushed towards the value $-1$ in contrast to the usual case 
for which it is difficult to go beyond $\omega _Q\approx -0.7$.  For 
$\Omega _{\rm m}\approx 0.3$, the model presented here predicts 
$\omega_Q\approx -0.82$. Finally, we establish the $\Omega _{\rm m}-\omega _Q$ 
relation for this model.


\newpage

Recent measurements of the relation between the luminous distance and the redshift 
using type Ia supernovae seem to suggest that our 
present Universe undergoes an accelerated expansion \cite{SNIa}. If 
confirmed, this means that our Universe 
is dominated by a type of matter with unusual properties. This matter would contribute 
by  $70 \%$ to the total energy of the Universe, the remaining $30 \%$ 
being essentially Cold Dark Matter ensuring that the Universe is spatially 
flat, $\Omega _0=1$, in agreement with the standard inflationary scenario. The 
unusual features of this fluid dominating the total energy of the Universe reveal 
themselves in the equation of state. One usually assumes that it takes the form 
$p=\omega _Q\rho $ leading to a negative  
$\omega _Q$. The cosmological constant ($\omega _Q=-1$) is a possible 
candidate but one has to face the task of explaining an energy scale of 
$\approx 5.7h^2 \times 10^{-47} \mbox{GeV}^4$, i.e. a value far from the natural 
scales of Particle Physics. Quintessence \cite{quint} is an alternative scenario with a 
homogeneous scalar field $Q$ whose equation of state is such 
that $-1 \le \omega _Q \le 0$. It has been shown \cite{quint} that, using
 the concept of ``tracking fields'', the coincidence and  the fine-tuning problems 
can be solved. The inverse power law potential 
$V(Q)=\Lambda ^{\alpha +4}/Q^{\alpha }$ is the prototype of such models which 
possess remarkable properties.
\par
We assume that the matter content of the Universe is composed of five 
different fluids: baryons, cold dark matter, photons, 
neutrinos and the quintessential field $Q$. The energy density of baryons and cold 
dark matter evolves as 
$\rho _{\rm m}=\rho _{\rm c} \Omega _{\rm m}(1+z)^3$ where $z$ is the 
redshift and $\rho _{\rm c}=8.1h^2\times 10^{-47}\mbox{GeV}^4$ the present 
value of the critical energy density. Observations 
indicate that $\Omega _{\rm m}=\Omega _{\rm b}+\Omega _{\rm cdm}
\approx 0.3$ \cite{KT,WCOS}. Photons and neutrinos have an energy 
density given by $\rho _{\rm r}=\rho _{\rm c} \Omega _{\rm r}(1+z)^4$. The 
contribution of radiation 
is negligible today since $\Omega _{\rm r}=\Omega _{\gamma }+\Omega _{\nu }
\approx 10^{-4}$. Finally, the fifth component is the scalar 
field $Q$. Its equation of state is characterized by $\omega _Q=
[\frac{1}{2}\dot{Q}^2-V(Q)]/[\frac{1}{2}\dot{Q}^2+V(Q)]$ where a dot represents 
a derivative with respect to cosmic time. A priori, 
$\omega _Q$ is not a constant and is such that $-1 \le \omega _Q \le 1$. Since 
the Universe is supposed to be spatially flat, we always have $\Omega _{\rm m}
+\Omega _{\rm r}+\Omega _Q=1$ which leads to $\Omega _Q\approx 0.7$. The inverse 
power law potential was first studied in Ref. \cite{RP}. If one 
requires that, during the radiation dominated era, the energy density 
of the scalar field is subdominant, i.e. $\rho _Q \ll \rho _{\rm r}$, and redshifts as 
$\rho _Q\propto a^{-4\alpha /(\alpha +2) }$ then one is automatically led to 
the inverse power law potential. This was the original motivation of Ref. \cite{RP} 
for considering this potential.  It is possible to 
find an exact solution to the Klein Gordon equation for which 
$Q\propto a^{4/(\alpha +2)}$. One can show that this solution 
is an attractor \cite{RP}. Then, if one follows the behaviour of the scalar 
field during the matter dominated era with the same potential, one can 
show \cite{RP} that $Q \propto a^{3/(\alpha +2)}$ is an exact solution 
which is still an attractor. For this solution, one has $\rho _Q \propto a^{-3\alpha 
/(\alpha +2)}$. The previous results are equivalent to say that the attractor is 
given by:
\begin{equation}
\label{attractor}
\frac{{\rm d}^2V(Q)}{{\rm d}Q^2}=\frac{9}{2}\frac{\alpha +1}{\alpha }(1-\omega _Q^2)H^2,
\end{equation}
during both the radiation and matter dominated epochs. We can re-write the parameter 
$\omega _Q$ as $\omega _Q=(\alpha \omega _{\rm B}-2)/(\alpha +2)$ where $\omega _{\rm B}$ 
is either $1/3$ or $0$. Since $\rho _Q$ 
redshifts slower than radiation or matter energy densities, the scalar field 
contribution becomes dominant at some stage of the evolution.
\par
As shown in Refs. \cite{quint}, this scenario possesses important 
advantages. Firstly, one 
can hope to avoid any fine-tuning. Indeed if the scalar field is on tracks today 
and begins to dominate and if, in addition, we require $\Omega _Q\approx 0.7$ 
then Eq. (\ref{attractor}) says that $Q\approx m_{\rm Pl}$ which implies 
that $\Lambda \approx (\Omega _Q\rho _{\rm c}m_{\rm Pl}^{\alpha })^{1/(4+\alpha )}
\approx 10^{10}\mbox{GeV}$ for $\alpha =11$, a very reasonable scale 
from the High Energy Physics point of view (we take $h=0.5$). Secondly, the solution will 
be on tracks today for a huge range of initial conditions including 
the equipartition for which $\Omega _{Qi}\approx 10^{-4}$. If one fixes the 
initial conditions at the end of inflation, i.e. $z\approx 10^{28}$, the 
allowed initial values for the 
energy density are such that $10^{-37}\mbox{GeV}^4 \stackrel{<}{\sim} \rho_Q 
\stackrel{<}{\sim}10^{61} \mbox{GeV}^4$ where $10^{-37}\mbox{GeV}^4$ is 
approximatively the background 
energy density at equality whereas $10^{61}\mbox{GeV}^4$ represents the 
background energy density at the initial redshift. If the scalar field starts at 
rest, this means that 
$1.8\times 10^{-10}m_{\rm Pl} \stackrel{<}{\sim} Q_{\rm i} \stackrel{<}{\sim}
0.16m_{\rm Pl}$ initially. Thirdly, the 
value of $\omega _Q$ is automatically such that $-1 \le \omega _Q \le 0$ 
today. For example, if $\alpha =11$ then one has $\omega _Q\approx -0.29$. This 
illustrates the fact that with inverse power law potentials, it is 
difficult to obtain values of $\omega _Q$ close to
$\omega _Q=-1$. This shortcoming can be partially removed if one 
considers smaller values of $\alpha $ or more general potentials of the form 
$V(Q)=\sum _kc_kQ^{-k}, k>0$\cite{quint}. However, it is not possible to reach 
a value lower than $\omega _Q\approx -0.7$. This seems in disagreement with recent 
estimates in Ref. \cite{Eft}. Finally, there exists a relation 
$\Omega _{\rm m}-\omega _Q$ which only depends on the functional form of the 
potential. This relation could also be used as an observational test of 
quintessence.
\par
This scenario raises the issue of the physical origin of the quintessential 
field. It is clear that this question should be adressed by the means of 
High Energy Physics \cite{Choi,MPR,PBM,R,MP}. Recently, an interesting 
supersymmetric (SUSY) model based upon the superpotential 
$W(Q)=\Lambda ^{3+a}/Q^a$, where $\Lambda \ll m_{\rm Pl}$, leading 
to an inverse power law scalar potential has been proposed 
\cite{PB}. This model suffers from serious problems. Firstly, it seems difficult to 
understand how, in this model, one can have $Q>\Lambda $ which is 
mandatory at the end of the evolution as $\Lambda$ is the UV cutoff at the 
gaugino condensation scale. Secondly, when the field is 
on tracks, one necessarily has $Q\approx m_{\rm Pl}$. One should 
take supergravity (SUGRA) into account. Let us emphasize 
that this is a generic property which comes from the very definition of 
a tracking solution. Therefore any model of quintessential tracking field 
coming from High Energy Physics must be based on SUGRA. In the context of 
the previous model, assuming that the K\"ahler potential is flat, $K(Q,Q^*)=Q^*Q$, 
one finds for the scalar potential:
\begin{equation}
\label{potcorrections}
V=e^{\frac{\kappa }{2}Q^2}\frac{\Lambda^{4+\alpha}}{Q^{\alpha}}
\biggl(\frac{(\alpha-2)^2}{4}-(\alpha+1)\frac{\kappa }{2}Q^2 +
\frac{\kappa^2}{4}Q^4\biggr),
\end{equation}
where $\kappa \equiv 8\pi G/c^4$ and $\alpha=2a+2$. This example is typical 
of the difficulties that one 
encounters in more general situations. A serious problem arises due to
the existence of negative contributions to the potential, a general 
property of any SUGRA model. These contributions entail that the energy density (and 
therefore the potential) becomes negative for $Q\approx m_{\rm Pl}$. We have studied 
numerically the case $\alpha =11$ in more details. The appearence of negative 
contributions depends on the value of $\Lambda $. For example, if 
$\Lambda \approx 8.7 \times 10^{10} \mbox{GeV}$, this occurs at $z\approx 2.24$. It  
is possible to avoid this problem by  changing the value of $\Lambda $. Indeed, there 
exist values of $\Lambda $ such that the negative contributions do not 
show up.  However, it is not 
possible to find a value such that $\Omega _Q\approx 0.7$. We have found that the 
best case is for $\Lambda \approx 2.1\times 10^{10}\mbox{GeV}$ for which 
$\Omega _Q\approx 0.055$ and $\omega _Q\approx -0.09$. This problem also 
occurs for other values of $\alpha $.
\par
The appearance of dangerous negative contributions to the potential is not 
accidental. The supergravity Lagrangian depends on two functions: the K\"ahler
potential $K(\phi _i,\phi _i^*)$ governing the kinetic terms of the boson fields and
the superpotential $W(\phi _i)$. The bosonic part of the 
Lagrangian can be derived from the following 
potential $G=\kappa K+ \ln(\kappa^3\vert W\vert^2)$. The kinetic terms are 
simply given by $K_{i \bar \j }\partial^{\mu}\phi^i\partial_{\mu}\bar\phi^{\bar \j }$ where  
$K_{i\bar \j}=\frac {\rm \partial }{\rm \partial\phi ^i}
\frac {\rm \partial } {\rm \partial\phi ^{\bar \j}}K$. The scalar potential is obtained as:
\begin{equation}
\label{defpotsugra}
V\equiv \frac{1}{\kappa^2}e^{G}(G^iG_i-3)+V_{\rm D},
\end{equation}
where $V_{\rm D}\ge 0$ is a term coming from the gauge sector. The negative 
contribution stems from the $-3$ term in the potential. The most 
natural way out is to impose that the superpotential 
vanishes and that the scalar potential is entirely due to a
non-flat K\"ahler potential. In this letter, we present a model where 
this can be achieved.
\par
Let us consider a supergravity model where there are two types of
fields, the quintessence field $Q$ and charged matter fields $(X,Y^i)$ under 
the gauge group. We assume that the gauge group of the model is broken along a flat
direction of the $D$ terms such that $X\ne 0, Y^i=0$ where $V_{\rm D}=0$. As already 
mentioned we impose that the scalar potential is positive to prevent any negative contribution 
to the energy density. This is achieved by considering that $\langle W \rangle =0$ when 
evaluated along the flat direction. Moreover we assume that
one of the gradients of the superpotential $W_Y$ does not vanish. The scalar 
potential is given by Eq. (\ref{defpotsugra}) evaluated along the 
flat direction and becomes
\begin{equation}
\label{potgeneral}
V=e^{\kappa K}K^{Y\bar Y}\vert W_Y\vert^2.
\end{equation}
As expected the scalar potential is positive and becomes a
function of the quintessence field $Q$ only. 
\par
As a guiding principle we now present a model which illustrates the 
quintessential property in SUGRA. We use string-inspired models with 
an anomalous $U(1)_X$ gauge symmetry \cite {witt}. 
We  consider the case of type I string theories\cite {anto}. The case of the usual 
compactification of the weakly coupled heterotic string\cite {lus} is phenomenologically 
disfavoured as the resulting scalar potential shows an exponential dependence 
on $Q$ \cite{us}. We suppose that the gauge group factorises as $G\times
U(1)_X$ where $G$ contains the standard model gauge group and
$U(1)_X$ is an anomalous Abelian symmetry. The fields of the
model split into three groups: the field $X$ has a charge $1$
under $U(1)_X$ and is neutral under $G$, the field $Y$ is a
matter field neutral under $G$ and of charge $-2$ under $U(1)_X$ 
while the matter fields $Y_i$ are charged under $G$ and possess 
charges $q_i\ne -2$ under $U(1)_X$. The matter fields $Y_i$ are spectators and 
will be discarded in the following.
We also assume that there is a modular symmetry $SL(2,Z)$ stemming from the
moduli space of the string compactification. We take into account a single modulus $t$ 
such that the radius of the compact manifold is $R_{\rm c}\equiv (t+\bar t) l_{\rm S}$ 
measured in units of the string scale $l_{\rm S}$. We assign different modular 
weights to the fields, i.e. $X$ is neutral, $Y$ has a weight $n_Y=n/p$ and $Q$ has 
a weight $n_Q=1-1/p$ where $n$ and $p$ are integers. Associated to $U(1)_X$ 
is the $D$ term potential:
\begin{equation}
\label{Dtermes}
V_{\rm D}=\frac{g_X^2}{2}\biggl(K^XX -2K^YY+\sum_i q_iK^{Y_i}Y_i
-\xi^2\biggr)^2,
\end{equation}
where $g_X$ is the $U(1)_X$ gauge coupling and  $\xi$ is a Fayet-Iliopoulos 
term. The K\"ahler potential of the effective supergravity theory describing 
the string theory at low energy is a function of the different fields $Q$, $X$ 
and $Y$ as well as the UV cutoff $m_{\rm c}=1/R_{\rm c}$. A compatible choice 
with the gauge and modular symmetries is
\begin{equation}
\label{kahler}
K=-\frac{1}{\kappa}\ln (t+\bar t) +XX^*+ \frac{(QQ^*)^p}{m_{\rm c}^{2p-2}}+\vert Y\vert^2 
\frac{(QQ^*)^{n}}{m_{\rm c}^{2n}}.
\end{equation}
The curvature along the flat direction possesses a delta function
singularity at the origin. The $D$ term potential
vanishes altogether along a flat direction where the field $X$ acquires
a vacuum expectation value breaking the Abelian 
symmetry $U(1)_X$ at $\langle X \rangle=\xi $, while the other fields 
vanish altogether. Expanding the superpotential in terms 
of Yukawa couplings, we obtain $W=\lambda (t)  X^2 Y +\dots $, where we have 
only taken into account the couplings such that 
$W_Y=\lambda(t) \langle X \rangle ^2 \neq 0 $ along the flat 
direction. The Yukawa coupling $\lambda (t)$ is a modular form of weight 
$-n/p-1$. Considering $Q=Q^*$, the scalar potential of this
supergravity model is given by
\begin{equation}
\label{potsugra}
V(Q)=\frac{\Lambda^{4+\alpha}}{Q^{\alpha}}e^{\frac{\kappa}{2}Q^2}.
\end{equation}
In this equation, the field $Q$ has been redefined such that the kinetic 
term takes the canonical form. The parameter $\Lambda $ can be expressed 
in terms of $m_{\rm c}$ according to $\Lambda^{4+\alpha}=
2^{\frac{\alpha }{2}}\lambda^2(t+\bar t)^{-1}\xi^4m_{\rm c}^{\alpha}$ where 
$\alpha=2n_Y$. In the type I string theories the Fayet-Iliopoulos term $\xi$ is 
moduli dependent \cite{iba}. Using the relation 
$m_{\rm Pl}^2\approx m_{\rm S}^8m_{\rm c}^{-6}$ for a 
six-dimensional compact space \cite {ark} and demanding $\lambda \approx 1$ to avoid 
any fine-tuning of the coupling constant, we obtain 
$\xi\approx (m_{\rm Pl}/m_{\rm c})^{\alpha/ 4} (\Omega_Q \rho_{\rm c})^{1/4}$. Imposing 
that the UV cutoff is $m_{\rm c}\approx 10^{14} \mbox{GeV}$ in order to be compatible 
with the fact that the evolution starts at the end of inflation and $\xi>10^2 \mbox{GeV}$ as 
the extra $U(1)_X$ symmetry has 
to be broken above the weak interaction scale, we deduce that
$\alpha \ge 11$ leading to $\Lambda \stackrel{>}{\sim} 10^{10}\mbox{GeV}$. The string 
scale is given by $m_{\rm S} \stackrel{>}{\sim} 10^{15}\mbox {GeV}$. The fine 
tuning problem can be avoided by allowing a sufficiently large 
value of $\alpha$. 
\par
Let us investigate the properties of the potential given by 
Eq. (\ref{potsugra}) for $\alpha=11$. First of all, the energy scale is 
now $\Lambda \approx 3.8 \times 10^{10} \mbox {GeV}$. Secondly, the 
presence of the exponential factor implies 
that this potential possesses arbitrary positive powers of $Q$. This 
could destroy the nice properties of the tracking solutions \cite{quint}. However, it 
is clear that this factor will play a role only at small redshifts since, initially, 
the value of the field is very small in comparison with the Planck mass. One expects 
a modification only at the very end of the evolution. We have checked 
numerically that the insensitivity to the initial conditions is totally preserved 
for the SUGRA potential. The initial value of $\rho _Q$ can change by $100$ orders of 
magnitude, the final result is always the same as in the usual 
case. The corresponding values for the quintessence field are now 
such that $5.8 \times 10^{-11}m_{\rm Pl} \stackrel{<}{\sim} Q_{\rm i}
\stackrel{<}{\sim}0.05 m_{\rm Pl}$. The evolution is displayed in Fig. 1 and is very 
similar to the evolution already found in Refs. \cite{quint}.
\begin{figure}
\begin{center}
\leavevmode
\hbox{%
\epsfxsize=9cm
\epsffile{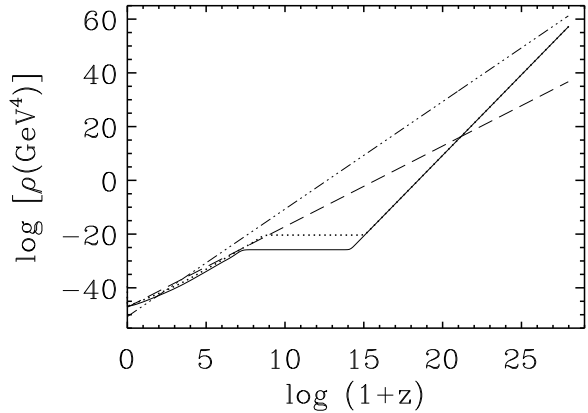}}
\end{center}
\caption{Evolution of the different energy densities. The 
dashed-dotted line represents the energy density of radiation whereas the 
dashed line represents the energy density of matter. The solid line is 
the energy density of quintessence in the SUGRA model with 
$\alpha =11$. The dotted line is the energy density of quintessence for the 
potential $V(Q)=\Lambda ^{4+\alpha }Q^{-\alpha }$ with the same $\alpha $. The 
initial conditions are such that equipartition, 
i.e. $\Omega _{Qi}\approx 10^{-4}$, is realized just after inflation.}
\label{nrj}
\end{figure}
Let us now study the evolution of the equation of state. The presence of the 
exponential factor is crucial. It has the effect of reinforcing the potential 
energy in comparison to the kinetic one and to push $\omega _Q$ 
towards the value $-1$ (recall that if the kinetic energy vanishes 
then $\omega _Q=-1$). This behaviour is illustrated in Fig. 2. For $\alpha =11$, the
final value of $\omega _Q$ is $-0.82$ whereas it is $\omega _Q\approx -0.29$ for the 
potential $V(Q)=\Lambda ^{4+\alpha }Q^{-\alpha }$. In this last case, the value 
of $\omega _Q$ 
changes from $\omega _Q\approx -0.63$ to $\omega _Q\approx -0.29$ when $\alpha $ goes from 
$2$ to $11$ at fixed $\Omega _Q=0.7$. In the SUGRA model, $\omega _Q$ 
changes from $\omega _Q\approx -0.89$ to $\omega _Q\approx -0.82$ in the same 
conditions; $\omega _Q$ does not strongly depend on $\alpha $. These 
properties render the SUGRA potential more attractive than the tracking solutions of 
Refs. \cite{quint} for which $w_Q$ cannot be less 
than $\approx -0.7$ (see Fig. 7 of that reference). In Ref. \cite{WCOS}, the constraint 
$-1\le \omega _Q \le -0.6$ is given whereas in Refs. \cite{Eft,PTW} a value between 
$-1$ and $-0.8$ is favoured. According to Ref. \cite{Eft}, $\omega _Q\approx -0.82$ 
is less than $1\sigma $ from the likelihood value. If the latter turns out to be 
confirmed the SUGRA model presented here could account for this.
\begin{figure}
\begin{center}
\leavevmode
\hbox{%
\epsfxsize=9cm
\epsffile{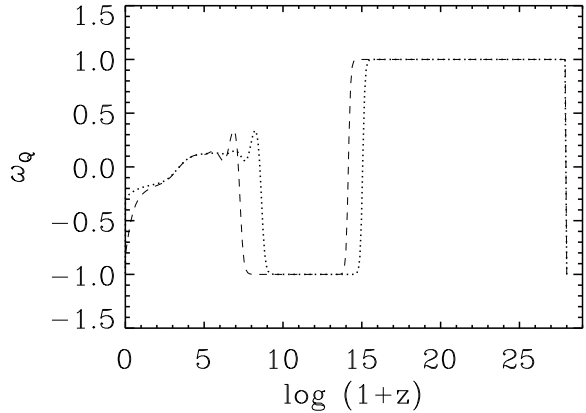}}
\end{center}
\caption{The dotted line represents the evolution of $\omega _Q$ for the potential 
$V(Q)=\Lambda ^{4+\alpha }Q^{-\alpha }$ with $\alpha =11$. The dashed 
line represents the evolution of $\omega _Q$ in the SUGRA model for the same value of 
$\alpha $. In this case $\omega _Q\approx -0.82$ today.}
\label{omega}
\end{figure}
Finally let us study the $\Omega _{\rm m}-\omega _Q$ relation which is displayed in Fig. 3.
\begin{figure}
\begin{center}
\leavevmode
\hbox{%
\epsfxsize=9cm
\epsffile{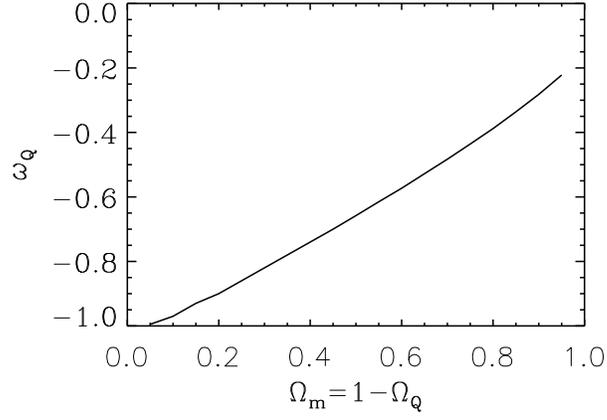}}
\end{center}
\caption{$\Omega _{\rm m}-\omega _Q$ relation for the SUGRA potential given by 
$V(Q)=\Lambda ^{4+\alpha }Q^{-\alpha }e^{\kappa Q^2/2}$ with $\alpha =11$.}
\label{oo}
\end{figure}
\begin{figure}
\begin{center}
\leavevmode
\hbox{%
\epsfxsize=9cm
\epsffile{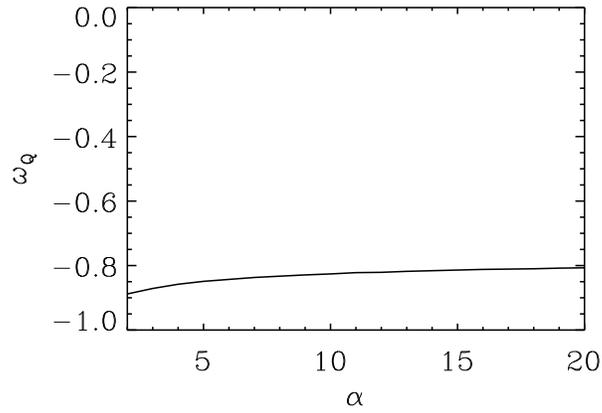}}
\end{center}
\caption{$\omega _Q-\alpha $ relation for the SUGRA potential.}
\label{alomega}
\end{figure}
The parameter $\omega _Q$ now varies between $-0.22$ and $-0.995$. If 
$\Omega _{\rm m}\approx 0.25$ instead of $0.3$ then 
$\omega _Q\approx -0.86$. The curve $\Omega _{\rm m}-\omega _Q$ has almost 
no dependence on $\alpha $. This is due to the fact that the value of $\omega _Q$ 
is mainly determined by the exponential factor which is $\alpha $ independent. In 
order to illustrate this property, the curve $\omega _Q-\alpha $ is displayed 
in Fig. 4. In this sense, the curve $\Omega _{\rm m}-\omega _Q$ presented here 
should be typical of any model based on SUGRA.
\par
In conclusion, we would like to emphasize the generic character of the 
results found in this letter. Nice properties arise (possibility to avoid the 
fine tuning problem, insensitivity 
to the initial conditions) if the quintessence field is on tracks today. This 
means that $Q\approx m_{\rm Pl}$ now and  SUGRA must be taken into account 
if one wishes to construct a realistic model. As a consequence, the scalar 
potential possesses an exponential factor which  pushes $\omega _Q$ 
towards the value $-1$. We have constructed explicitly a natural model in 
this context. If $\alpha \ge 11$ then the fine tuning problem can be overcome. In this 
simple case, we find that $\omega _Q\approx -0.82$ if $\Omega _{\rm m}\approx 0.3$.
\par
Finally let us comment on the supersymmetry breaking issue. When SUGRA is broken 
by the dilatonic $F$ term the model 
is not modified whereas in the case where it is broken by the moduli $F$ 
term the coupling between $Q$ and $t$ induces an inverse power law 
potential. More studies are required in the moduli breaking scenario. This 
question will be adressed elsewhere \cite{us2}.

\vspace{0.5cm}
\noindent {\bf \large Acknowledgements:}
It is a pleasure to thank Martin Lemoine for useful exchanges and comments 
and for his invaluable help in the writing of the codes used in this 
paper. We would like to thank Alain Riazuelo for useful discussions.

\end{document}